\documentclass[nofootinbib,notitlepage,amssymb, nobibnotes, aps, prd,showpacs,showkeys]{revtex4-1}

\usepackage{empheq}
\usepackage{hyperref}
\usepackage{color}

\newcommand{\ie}{\emph{i.e.}}

\usepackage{amsmath}
\usepackage{amssymb}

\begin{document}

\title{Has the Goldstone theorem been revisited?}
\author{A. Guerrieri$^\#$, A.~Pilloni$^{*}$}
\affiliation{\mbox{$^\#$Dipartimento di Fisica and INFN, Universit\`a di Roma `Tor Vergata',}\\ Via della Ricerca Scientifica 1, I-00133 Roma, Italy\\
\mbox{$^*$Dipartimento di Fisica and INFN, `Sapienza' Universit\`a di Roma,}\\ P.le Aldo Moro 5, I-00185 Roma, Italy
}

\begin{abstract}
A recent paper (arXiv:1404.5619) claimed the presence of a loophole in the current-algebra proof of Goldstone Theorem. The enforcing of manifest covariance would lead to contradictory results also in scalar theory. We show that the argument proposed is not in contradiction with covariance, thus not invalidating the theorem. Moreover, the counterexample proposed of a scalar operator with a non-zero vacuum expectation value in an unbroken theory is ill-defined.
\end{abstract}
\pacs{11.10.-z, 11.30.Qc}
\keywords{Goldstone theorem, Goldstone bosons, Spontaneous symmetry breaking}
\maketitle

In this brief note we point out some remarks about a recently appeared paper by A.~Kartavtsev~\cite{Kartavtsev} which reports a loophole in the proof of the Goldstone theorem~\cite{Goldstone:1961eq,Goldstone:1962es}. According to the author of Ref.~\cite{Kartavtsev}, it is uncorrect to assume manifest covariance of correlators. Since this is required by the usual current-algebra proof, the theorem is flawed. For instance, also simple cases as scalar free theory would provide non-covariant spectral density, thus evading the theorem.

We briefly review Kartavtsev's argument. Let us consider a charged scalar theory, and the vacuum expectation value of
\begin{equation}
 \left\langle 0 \right| O(0) J^{\mu}(x)\left|0\right\rangle=\sum_n \left\langle 0 \right| O(0) \left|n\right\rangle \left\langle n \right|J^{\mu}(0)\left|0\right\rangle e^{ip_n x}\,,
\label{pippone}
 \end{equation}
where $O$ is a suitable scalar local operator and $J^{\mu}$ is the $U(1)$ conserved current $J^{\mu}=-i(\phi^{\dag}\partial^{\mu}\phi - \phi\partial^{\mu}\phi^{\dag})$.

The spectral density function, defined as
\begin{equation}
 \rho^{\mu}(P)=\sum_n \left\langle 0 \right| O(0) \left|n\right\rangle \left\langle n \right|J^{\mu}(0)\left|0\right\rangle \delta^4(P-p_n)\,,
\label{pippone2}
 \end{equation}
can be expressed in terms of a Lorentz invariant function 
$\rho^{\mu}(P)=P^{\mu}\rho(P^2)\,\theta(P^0)/(2\pi)^3$. 
A key point in the proof of the Goldstone theorem is that, being $P_{\mu}J^{\mu}=0$, it must be 
\begin{equation}
 \rho(P^2)=c\,\delta(P^2),
\label{keycond}
 \end{equation}
thus showing the presence of a massless state if $c\neq0$.
In \cite{Kartavtsev}, it is argued that $\rho^{\mu}(P^2)$ could be orthogonal to $P^{\mu}$ by itself and automatically enforce current conservation, thus circumventing the condition \eqref{keycond}. An example of this is the free field theory.
We can write eq.~\eqref{pippone2} as~\footnote{We use the covariant normalization for the states.}
\begin{equation}
 \rho^{\mu}(P)=\int \frac{d^3 q_1}{(2\pi)^3 2\omega_1} \frac{d^3 q_2}{(2\pi)^3 2\omega_2} \left\langle 0 \right| O(0) \left|q_1 q_2\right\rangle \left\langle q_1 q_2 \right|J^{\mu}(0)\left|0\right\rangle \delta^4(P-q_1-q_2)\,,
\label{integralozzo}
 \end{equation}
and in canonical formalism, it holds
\begin{multline}
\label{JnuFreeFields}
J^\mu(x)=\int\frac{d^3q_1}{(2\pi)^3 2\omega_1}
\frac{d^3q_2}{(2\pi)^3 2\omega_2}
\Bigg[(q_1+q_2)^\mu \left(-e^{i(q_1-q_2)x} a^\dagger_{q_1} a_{q_2} + e^{i(q_1-q_2)x}b^\dagger_{q_1} b_{q_2}\right)\\
+(q_1-q_2)^\mu \left(e^{-i(q_1+q_2)x}b_{q_1} a_{q_2} + e^{i(q_1+q_2)x}b^\dagger_{q_1}a^\dagger_{q_2}\right)\Bigg]\,.
\end{multline}

In free theory, the spectral density in eq.~\eqref{pippone} is supported over two-particle states only, and
\begin{equation}
 \left\langle q_1 q_2 \right|J^\mu(x)\left|0\right\rangle = (q_1 - q_2)^{\mu}\,e^{i(q_1+q_2)x}\,.
\end{equation}
Current conservation is fulfilled: $\partial_{\mu}\left\langle q_1 q_2 \right|J^{\mu}(x)\left|0\right\rangle=i(q_1+q_2)\cdot(q_1 - q_2)=i(m^2-m^2)=0$. This would imply $P_{\mu}\rho^{\mu}(P)=0$, \ie\ $\rho^{\mu}$ is \emph{orthogonal} to $P^{\mu}$. On the other hand, manifest covariance would require $\rho^{\mu} \propto P^\mu$, \ie\ spectral density should be \emph{parallel} to the momentum. The author of Ref.~\cite{Kartavtsev} concludes that the two conditions are incompatible, and the enforcing of covariance would lead to erroneous conclusions, like the Goldstone theorem.

This last point should be enough to show that Kartavtsev's conclusion is wrong:
if $\rho^{\mu}$ is at the same time orthogonal and parallel to $P^{\mu}$, it must be $\rho^{\mu}(P)\equiv 0$. This is indeed compulsory in free theory, the $U(1)$ symmetry being unbroken. The charge must annihilate the vacuum, so that 
\begin{equation}
 0\equiv \left\langle 0 \right| O(0) Q\left|0\right\rangle=\int d^3 x \left\langle 0 \right| O(0) J^0(x)\left|0\right\rangle=\int d^3 x \int d^4P e^{iPx} \rho^0(P)=\int dP^0 e^{iP^0x^0} \rho^0(P^0,\vec{0})\,.
\end{equation}

On the other hand, we cannot build a covariant projector automatically orthogonal to $P^\mu$ with one Lorentz index: the only possibility would be
\begin{equation}
\rho^\mu(P) = f(P^2) \left( P^\mu - g^{0\mu} \frac{P^2}{P^0}\right)\,, 
\end{equation}
which would be manifestly non-covariant. Instead, the explicit covariance of the spectral density comes from the covariance of the current and the invariance of the vacuum:
\begin{equation}
\left\langle0|J^\mu\!\left(x\right) O(0) |0\right\rangle = \left\langle0|U^\dagger\left(\Lambda\right)J^\mu\!\left(x\right) U\left(\Lambda\right)O(0) |0\right\rangle = {\Lambda^\mu}_\nu \left\langle0|J^\nu\!\left(\Lambda^{-1} x\right) O(0) |0\right\rangle\,,
\end{equation}
and consequently $\rho^\mu\!\left(P\right) = {\Lambda^\mu}_\nu\, \rho^\nu\left(\Lambda^{-1} P\right)$, which implies $\rho^\mu\!\left(P\right) \propto P^\mu$. We can thus evaluate the integral in the center-of-mass frame, where it reads
\begin{equation}
 \rho^{\mu}_\text{com} = \int \frac{d^3 q_1}{(2\pi)^3 2\omega_1}\frac{d^3 q_2}{(2\pi)^3 2\omega_2} \left\langle 0 \right| O(0) \left|q_1 q_2 \right\rangle (q_1-q_2)^{\mu} \,\delta(P^0 - \omega_1 - \omega_2)\, \delta^3(q_1 + q_2) = 0\,.
\end{equation}

However, since manifest covariance is contested in Ref.~\cite{Kartavtsev}, the explicit calculation in a generic frame reported in next section could be more convincing.

Another argument discussed in Ref.~\cite{Kartavtsev} deals with the polar representation of the complex scalar field. We choose the order parameter to be
\begin{equation}
 O(x) = \theta(x) = \arctan \frac{\Im \phi}{\Re \phi}\,.
\end{equation}
The spectral density is non-vanishing because
\begin{equation}
 \left[\theta(x),J^0(y)\right]_{x^0=y^0} \equiv i \delta^3 (\vec x - \vec y)
 \label{theta}
\end{equation}

Kartavtsev asserts that a massless state must appear according to Goldstone theorem for any symmetric form of the potential, in particular for the (unbroken) free theory. Since no massless excitation appear in free theory, the theorem is challenged. Unfortunately, the polar representation is singular in free theory: if $\rho = 0$, $\theta$ is undefined, as we see in the definition when $\Im\phi,\Re\phi \to 0$. Moreover, quantum fluctuations around zero are not well described by the positive-definite field $\rho$. On the other hand, the representation is meaningful when $\rho$ fluctuates around a non-zero value only, \ie\ when the symmetry is spontaneously broken. This explains why the commutator in~\eqref{theta} is identically non-zero. We recall that spontaneous symmetry breaking can actually occur in a free theory: in a massless real scalar theory, the shift symmetry $\phi \to \phi + c$ is spontaneously broken. As expected by Goldstone theorem, a massless boson must appear, and 
it is the $\phi$ itself.

To conclude, the arguments presented in Ref.~\cite{Kartavtsev} are flawed. Free theory does not provide any counterexample to the Goldstone theorem, nor to general covariance.

{\bf \emph{Acknowledgements.}} We whish to thank Massimo Testa for useful discussions. 
 
\section*{Spectral density of free theory}
\label{sec:appendix}
We evaluate the spectral density of free theory in a generic frame. We start from the definition in eq.~\eqref{integralozzo}:
\begin{equation}
 \rho^{\mu}(P)=\int \frac{d^3 q_1}{(2\pi)^3 2\omega_1} \frac{d^3 q_2}{(2\pi)^3 2\omega_2} \left\langle 0 \right| O(0) \left|q_1 q_2\right\rangle \left\langle q_1 q_2 \right|J^{\mu}(0)\left|0\right\rangle \delta^4(P-q_1-q_2)\,.
\end{equation}

As already shown, in free theory we have $\left\langle q_1 q_2 \right|J_{\mu}(0)\left|0\right\rangle = (q_1 - q_2)_{\mu}$.
Moreover 
\begin{equation}
 \left\langle 0 \right| O(0) \left|q_1q_2\right\rangle=F((q_1+q_2)^2,(q_1-q_2)^2,(q_1+q_2)\cdot(q_1-q_2))\equiv F((q_1+q_2)^2)\,,
\end{equation}
being $(q_1-q_2)^2=4m^2 - (q_1+q_2)^2$, and $(q_1+q_2)\cdot(q_1-q_2)=0$. We see that our result does not depend on the details of the order parameter, nor whether it has a non-zero vacuum expectation value.

Substituting $ d^3q_2/2\omega_{2}=d^4q \,\delta(q_2^2-m^2)\,\theta(q_2^0)$ and integrating the 4D-delta function 
\begin{equation}
 \rho^{\mu}(P)= F(P^2) \int \frac{d^3 q_1}{(2\pi)^3 2\omega_1} (2q_1-P)^{\mu} \delta((P-q_1)^2-m^2)\,\theta(P^0-q_1^0)\,.
\end{equation}

Without loss of generality, we can assume that $P^{\mu}=(P^0,0,0,P^z)^{\mu}$. The delta function
\begin{equation}
 \delta((P-q_1)^2-m^2)=\delta(P^2-2P^0\omega_1 + 2P^z|q_1|\cos\theta)=\frac{1}{2P^z|q_1|} \delta\left(\cos\theta-\frac{2P^0\omega_1-P^2}{2P^z|q_1|}\right)\,.
\end{equation}

Finally, $d^3 q_1=|q_1|\omega_1d\omega_1d\cos\theta d\phi$, and
\begin{equation}
 \rho^{\mu}(P)=F(P^2) \int \frac{d\omega_1d\cos\theta d\phi}{(2\pi)^3 4P^z} (2q_1-P)^{\mu} \delta\left(\cos\theta-\frac{2P^0\omega_1-P^2}{2P^z|q_1|}\right)\theta(P^0-\omega_1)\,.
\end{equation}

The theta function and the on-shell condition would restrict the domain to $m \le \omega_1 \le P_0$, but the enforcing of $-1 \le \cos\theta \le 1$ leads to the stronger inequality
\begin{equation}
\lambda_- = \frac12\left(P^0 - P^z \sqrt{1-\frac{4m^2}{P^2}}\right) \le \omega_1 \le \frac12\left(P^0 + P^z \sqrt{1-\frac{4m^2}{P^2}}\right) = \lambda_+\,.
\end{equation}

The support of the spectral function turns out to be $P^2 > 4m^2$ and $P^0 > 0$.
If we consider 
\begin{equation}
 (2q_1-P)^{\mu}=\left( 2\omega_1 - P^0, 2|q_1|\sin\theta \cos\phi, 2|q_1|\sin\theta \sin\phi,2|q_1|\cos\theta-P^z\right)^{\mu}\,,
\end{equation}
it is trivial that the spatial transverse components vanish upon $d\phi$ integration.

As for the time component, 
\begin{multline}
 \rho^{0}(P)=2\pi F(P^2) \int_{\lambda_-}^{\lambda_+} \frac{d\omega_1}{(2\pi)^3 4P^z} (2\omega_1-P^0) = \frac{F(P^2)}{(2\pi)^2 4P^z}\int_{\lambda_-}^{\lambda_+} d\omega_1 (2\omega_1-P^0)\\
 =  \frac{F(P^2)}{(2\pi)^2 4P^z} \left[\lambda_+^2 - \lambda_-^2 - P^0 \left(\lambda_+ - \lambda_-\right)\right]=  \frac{F(P^2)}{(2\pi)^2 4P^z} \left[P^0 P^z\sqrt{1-\frac{4m^2}{P^2}} - P^0 P^z\sqrt{1-\frac{4m^2}{P^2}}\right]=0\,,
\end{multline}
and the same happens for the longitudinal component:
\begin{multline}
 \rho^{z}(P)=2\pi F(P^2) \int_{\lambda_-}^{\lambda_+} \frac{d\omega_1}{(2\pi)^3 4P^z} \left(2|q_1| \frac{2P^0\omega_1-P^2}{2P^z|q_1|} -P^z\right) \\
 =\frac{F(P^2)}{(2\pi)^2 4(P^z)^2} \int_{\lambda_-}^{\lambda_+} d\omega_1 (2P^0\omega_1-(P^0)^2+(P^z)^2-(P^z)^2)\\=
 \frac{F(P^2) P^0}{(2\pi)^2 4(P^z)^2} \left[\lambda_+^2 - \lambda_-^2 -P^0\left(\lambda_+ - \lambda_-\right)\right]=0\,.
\end{multline}

As expected by manifest covariance, the correlator vanishes in any frame.

\end{document}